
\magnification=\magstep 1
\baselineskip=20 pt
\font\bigbf = cmbx10 scaled\magstep 1
 1
 1

\line{}

\noindent

\centerline{\bigbf \hfill Instantons and the spectral function of \hfill}
\centerline{\bigbf \hfill electrons in the half-filled Landau level \hfill}

\vskip 1.0cm

\centerline{\hfill Yong Baek Kim and Xiao-Gang Wen \hfill}
\centerline{\hfill \it Department of Physics, Massachusetts Institute of
Technology \hfill}
\centerline{\hfill \it Cambridge, Massachusetts 02139 \hfill}

\vskip 1.0cm

\centerline{\hfill ABSTRACT \hfill}

\vskip 0.3cm

\midinsert
\narrower
{\noindent\tenrm

We calculate the instanton-anti-instanton action $S_{M {\bar M}} (\tau)$
in the gauge theory of the half-filled Landau level.
It is found that $S_{M {\bar M}} (\tau) = (3 - \eta) \left [ \Omega_0 (\eta)
\ \tau \right ]^{1 / (3 - \eta)}$ for a class of interactions $v ({\bf q}) =
V_0 / q^{\eta} \ ( 0 \leq \eta < 2 )$ between electrons.
This means that the instanton-anti-instanton pairs are
confining so that a well defined `charged' composite fermion can exist.
It is also shown that $S_{M {\bar M}} (\tau)$ can be used to calculate
the spectral function of electrons from the microscopic theory
within a semiclassical approximation. The resulting spectral function
varies as
$e^{ - \left [ \Omega_0 (\eta) / \omega \right ]^{1 / ( 2 - \eta ) } }$
at low energies.

\vskip 0.2cm
\noindent
PACS numbers: 73.40.Hm, 71.27.+a, 11.15.-q
}
\endinsert

\vfill\vfill\vfill
\break

\vskip 0.5cm

The idea of composite fermions [1] has been used to explain the hierarchical
structure of fractional quantum Hall (FQH) liquids.
Using the fact that composite fermions in the filling fraction
$\nu = 1/2$ state feel
zero magnetic field at the mean field level and employing the fermionic
Chern-Simons gauge theory of the FQH states [2,3],
Halperin, Lee and Read [4]
constructed a renormalized Fermi liquid theory of the half-filled
Landau level.
As a direct consequence of a well defined Fermi surface at the mean field
level, the integral quantum Hall effect of composite fermions can be
viewed as an extreme form of the Shubnikov-de Haas effect [4].
Although several experiments [5-8] already appeared
to support the existence of a well defined Fermi surface, from the theoretical
point of view, the strong gauge field fluctuations and the resulting divergent
effective mass of fermion reflect the difficulty of explaining the success of
the mean field theory.

It is well known that, in (2+1) dimensional compact Maxwell
U(1) gauge theory or QED, the existence of the instanton solutions leads
to the confinement of charges and significantly changes the infrared
behavior of the theory [9].
This happens because instantons or monopoles
effectively change the logarithmic interaction to a linear one.

Thus one may worry about the confinement of composite fermions
in the compact gauge theory of the $\nu = 1/2$ state.
By calculating instanton-anti-instanton action, it is shown that
instantons are confining in the gauge theory of the $\nu = 1/2$ state
so that well defined composite fermions can
exist. This problem is important because, if instantons were not confining,
well-defined composite fermions would not exist and there would be no
well-defined Fermi surface for composite fermions, which is necessary to
explain experiments.

We also calculated one electron Green's function of the $\nu = 1/2$
state from the microscopic theory.
Here we are in a completely new situation (compared to the usual case)
that the electron operator
not only creates a composite fermion but also creates flux quanta.
Therefore we need to develop a new method to calculate the correlation
functions of electrons.
It is found that, using the calculated instanton-anti-instanton action,
one can compute the spectral function of electrons in a semiclassical
approximation. The resulting spectral function shows a strong suppression
at low energies which may explain a recent measurement of the low
temperature I-V tunneling characteristics
of a double-layer FQH system near $\nu = 1/2$ [10].
It is worthwhile to mention that this highly suppressed spectral density
in the infrared limit is not realized in the usual Fermi liquid.
There are some numerical calculations of small size systems [11,12]
and a phenomenological model based on the low-lying density fluctuations [11]
successfully explained the experiment.
However, the calculation presented here is a microscopic derivation
which shows some deviations from the phenomenological construction.

In a recent paper (see also Refs.[13,14]),
Diamantini, Sodano and Trugenberger [15] discussed the instanton effect
in a (2+1) dimensional compact U(1) gauge theory with the Chern-Simons term.
It was found that the effect of the Chern-Simons term dominates the role
of monopoles in the infrared limit so that the monopoles are linearly
confining.
Our problem is more delicate because there are also fermions in the
theory and this fermionic degree of freedom generates particle-hole
excitations across the Fermi surface which may affect the dynamics
of the gauge field [16,17].
Recently, Nagaosa [17] investigated a dissipative U(1) gauge model
that is a simplified version of the gauge theory of high $T_c$
superconductors.
He found that the above mentioned low energy excitations give
rise to a dissipative effect on the gauge field so that the confinement
of charges is strongly suppressed [17].

In the gauge theory of the $\nu = 1/2$ state, both of the low energy
particle-hole excitations and the Chern-Simons term exist.
In this paper, it is shown that the effect of the low energy
particle-hole excitations dominates the effect of
the Chern-Simons term for a class of interactions between
electrons:
$v({\bf q}) = V_0 / q^{\eta} \ ( 0 \leq \eta < 2 )$.
This leads to a confinement of instantons similar to that in
the gauge theory of high $T_c$ superconductors [17].

In the fermionic Chern-Simons gauge theory of FQH states [2-4],
the problem of interacting electrons in a uniform magnetic field can
be transformed to an equivalent system in which a fermion is minimally
coupled to a
statistical gauge field $a_{\mu} ({\bf r})$ as well as the uniform
magnetic field.
The fermion operator $\psi^{\dagger}$ is related to the electron operator
$\psi^{\dagger}_e$ [2,4] as
$$
\psi^{\dagger}_e = \psi^{\dagger} \ \exp \left [ i \ {\tilde \phi} \int \ d^2
r'
\ \varphi ({\bf r}-{\bf r'}) \ \rho ({\bf r'}) \right ] \ ,
\eqno{(1)}
$$
where $\varphi ({\bf r}-{\bf r'})$ is the angle between ${\bf r} - {\bf r'}$
and the $x$ axis, and $\rho ({\bf r})$ is the electron or fermion density
operator at the point ${\bf r}$.
For fermionic theory, ${\tilde \phi}$ should be even integers
and especially ${\tilde \phi} = 2$ for $\nu = 1/2$ state.
In this fermionic language, the Hamiltonian can be written as [4]
$$
\eqalign{ H &= H_0 + V  \ , \cr
H_0 &= {1 \over 2 m^*} \int \ d^2 r \ \psi^{\dagger}
(-i \nabla - \Delta {\bf a})^2
\psi \ , \cr
V &= {1 \over 2} \int \ d^2 r \ d^2 r' \ v({\bf r} - {\bf r'})
:\rho ({\bf r}) \rho ({\bf r'}): \ , }
\eqno{(2)}
$$
where the colons represent the normal ordering and $m^*$ is the effective
mass of the fermion.
Here we assume that interaction between electrons takes a form (in
Fourier space):
$v({\bf q}) = V_0 / q^{\eta} \ (0 \leq \eta < 2)$.
$\Delta {\bf a} ({\bf r}) = {\bf a} ({\bf r}) - e {\bf A} ({\bf r})$
is the fluctuation above
the mean field configuration ${\bf a} ({\bf r}) = e {\bf A} ({\bf r})$ and
${\bf a} ({\bf r}) = {\tilde \phi} \ \int \ d^2 r' \ \nabla
\varphi ({\bf r}-{\bf r'}) \ \rho ({\bf r'})$.

In the rest of the paper, $a_{\mu} ({\bf r})$ means the fluctuation above
the mean field configuration, {\it i.e.}, $\Delta a_{\mu} ({\bf r})$.
We will use the Euclidean functional-integral formalism and choose the
temporal gauge in which $a_0 = 0$.
The effective action of the gauge field can be obtained after integrating
out the fermions in the original action.
Since only the transverse fluctuation of the gauge field is important
in the low energy limit [16,17], we will drop the longitudinal fluctuation
from now on. It turns out that the gauge field propagator is not
renormalized by the fluctuations beyond the random phase approximation [18].
Therefore we can employ the same gauge field fluctuation as that of
Ref [4].
The effective gauge field action can be written as the following [16,17].
$$
\eqalign{ S_{\rm eff} &= S_0 + S_{\rm cs} \ , \cr
S_0 &= {1 \over 2} \int \ {d^2 q \over (2 \pi)^2} \ {d\omega \over 2\pi}
\Bigl [ \ \varepsilon ({\bf q}, \omega) e_{\alpha} ({\bf q}, \omega)
e_{\alpha} (-{\bf q}, -\omega) +
\mu ({\bf q}, \omega) b({\bf q}, \omega) b(-{\bf q}, -\omega)
\ \Bigr ] \ , \cr
S_{\rm cs} &= - i \int \ d{\tau} \ d^2 r \
{m \over 4} \ \epsilon_{\mu \nu \lambda} \ a_{\mu} \ f_{\nu \lambda} \ ,
}
\eqno{(3)}
$$
where $e_{\alpha} = \partial_0 a_{\alpha}$ $(\alpha = 1, 2)$,
$b = \partial_1 a_2 - \partial_2 a_1$, $m = 1 / (2 \pi {\tilde \phi})$,
and $f_{\mu \nu} = \partial_{\mu} a_{\nu} - \partial_{\nu} a_{\mu}$.
The dielectric function $\varepsilon ({\bf q}, \omega)$ and
the magnetic permeability $\mu ({\bf q}, \omega)$ (for the statistical
gauge field) are given by
$\varepsilon ({\bf q}, \omega) = {\nu_1 \over |\omega| q}$
$( \nu_1 = {2 n_e \over m^* v_F} )$ and
$\mu ({\bf q}, \omega) = {1 \over 12 \pi m^*} +
{v({\bf q}) \over (2 \pi {\tilde \phi})^2}$.

Before the calculation of instanton-anti-instanton action and
the demonstration of confinement of instantons, we would like to show
a relation between the electron Green's function and
the instanton-anti-instanton action.
The one electron Green's function $G_{+} (\tau) =
\langle \psi_e ({\bf 0}, \tau) \psi^{\dagger}_e ({\bf 0}, 0) \rangle$ can
be calculated semiclassically in the spirit of the WKB approximation [19].
In the functional integral approach, $G_{+} (\tau)$ can be written as
$$
\eqalign{
G_{+} (\tau) &= \int \ D \psi^{\dagger} \ D \psi \ D a_{\mu} \
\psi_e (\tau) \ \psi^{\dagger}_e (0) \
e^{ - S (\psi^{\dagger}, \psi, a_{\mu}) } \cr
&= \int \ D \psi^{\dagger} \ D \psi \ D a_{\mu} \
\psi (\tau) \ \psi^{\dagger} (0) \ \delta (M {\bar M}) \
e^{ - S (\psi^{\dagger}, \psi, a_{\mu}) } \ ,
}
\eqno{(4)}
$$
where $S (\psi^{\dagger}, \psi, a_{\mu})$ is the action given by Eq.(2).
Notice that, since an electron is a fermion plus two  flux quanta,
the creation and annihilation of electrons at times zero and
$\tau$ not only create and annihilate a fermion, it also create
and annihilate flux quanta. The creation and annihilation
of flux quanta is represented by
a singular boundary condition on
the gauge field and $\delta (M {\bar M})$ represents this
boundary condition. Notice that the creation (annihilation)
of a flux quantum corresponds to inserting a (anti-)monopole in space-time.
Formally, Eq.(4) can be rewritten as
$$
G_{+} (\tau) = \int \ D a_{\mu} \ \delta (M {\bar M}) \
\langle \psi (\tau) \ \psi^{\dagger} (0) \rangle_{a} \
e^{ - S_{\rm eff} (a_{\mu}) } \ ,
\eqno{(5)}
$$
where
$$
\eqalign{
\langle \psi (\tau) \ \psi^{\dagger} (0) \rangle_{a} &=
\int \ D \psi^{\dagger} \ D \psi \
\psi (\tau) \ \psi^{\dagger} (0)
\ e^{ - S (\psi^{\dagger}, \psi, a_{\mu}) }
/ e^{ - S_{\rm eff} (a_{\mu}) } \ , \cr
e^{ - S_{\rm eff} (a_{\mu}) } &= \int \ D \psi^{\dagger} \ D \psi
\ e^{ - S (\psi^{\dagger}, \psi, a_{\mu}) } \ .
}
\eqno{(6)}
$$
Notice that both of $\langle \psi (\tau) \ \psi^{\dagger} (0) \rangle_{a}$
and $e^{ - S_{\rm eff} (a_{\mu}) }$ are not gauge invariant.
Let us introduce $S_{\rm eff} ( a_{\mu}, j_{\mu} ) =
S_{\rm eff} ( a_\mu ) - \int \ d^3 r \ a_{\mu} \ j_{\mu}$, where
$j_{\mu}$ is the fermion current corresponding to the straight line path
and has the following form.
$$
j_{\mu} = \left [ \theta (x_0 - \tau) - \theta (x_0) \right ]
\delta (x_1) \delta (x_2) \delta_{\mu 0} \ .
\eqno{(7)}
$$
Now we can write the integrand of the functional integral as a product
of two gauge invariant objects:
$$
G_{+} (\tau) = \int \ D a_{\mu} \ \delta (M {\bar M}) \
\left [ \ \langle \psi (\tau) \ \psi^{\dagger} (0) \rangle_{a}
\ e^{ - \int d^3 r \ a_{\mu} j_{\mu} } \ \right ] \
e^{ - S_{\rm eff} (a_{\mu}, j_{\mu}) } \ ,
\eqno{(8)}
$$
where $\langle \psi (\tau) \ \psi^{\dagger} (0) \rangle_{a}
\ e^{ - \int d^3 r \ a_{\mu} j_{\mu} }$ and
$e^{ - S_{\rm eff} (a_{\mu}, j_{\mu}) }$ are gauge invariant.
Notice that the two terms in $S_{\rm eff} ( a_{\mu}, j_{\mu} )$
are not gauge invariant respectively, due to the presence of
the monopoles for the first term and non-conservation of the current
$j_\mu$ for the second term (an electron is created and annihilated).
However, the total effective action $S_{\rm eff} ( a_{\mu}, j_{\mu} )$
is gauge invariant.
Notice also that, in the semiclassical limit, the paths of the fermions are
close to the straight line path in a given gauge field background,
thus the factor $e^{ - \int d^3 r \ a_{\mu} j_{\mu} }$ is almost
compensated by the contribution from the fermions
$\langle \psi (\tau) \ \psi^{\dagger} (0) \rangle_{a}$.
Therefore, the saddle point of the integrand is dominated by
$e^{ - S_{\rm eff} (a_{\mu}, j_{\mu}) }$.
By taking out the saddle point value
$S_{\rm eff} ({\bar a}_{\mu}, j_{\mu})$ in which the boundary condition
$\delta (M {\bar M})$ should be incorporated, one can do the following
semiclassical approximation
$$
G_{+} (\tau) \approx e^{ - S_{\rm eff} ({\bar a}_{\mu}, j_{\mu}) }
\ \int \ D \delta a_{\mu} \
\langle \psi (\tau) \ \psi^{\dagger} (0) \rangle_{a}
\ e^{ - \int d^3 r \ a_{\mu} j_{\mu} }
\ e^{ - S_{\rm eff} (\delta a_{\mu}) } \ ,
\eqno{(9)}
$$
where $\delta a_{\mu}$ is the fluctuation around the saddle point and
$S_{\rm eff} (\delta a_{\mu})$ can be taken as a quadratic expansion in
$\delta a_{\mu}$.
Combined with the boundary condition on the gauge field, $j_{\mu}$ of
Eq.(7) is exactly the source of the instanton--anti-instanton
(or monopole-antimonopole) solution of the effective gauge field
action [15].
In other words, the monopole and antimonopole are connected by a string of
source $j_{\mu}$ in Euclidean space.
{}From these arguments, we can identify $S_{\rm eff} ({\bar a}_{\mu}, j_{\mu})$
as the monopole-antimonopole action $S_{M {\bar M}}$.
Therefore, the electron Green's function can be written as
$$
G_{+} (\tau) \approx G_0 (\tau) \ e^{ - S_{M {\bar M}} (\tau) } \ ,
\eqno{(10)}
$$
where $G_0 (\tau)$ is at most an algebraically decaying function of
$\tau$ [20] because of the above mentioned compensation effect
in the semiclassical approximation. It will be shown that
$e^{ - S_{M {\bar M}} (\tau) }$ is the dominant suppression factor
of low energy electron spectral function.

Now let us concentrate on the evaluation of $S_{M {\bar M}} (\tau)$
in our model.
As a simpler case, we can consider the compact Maxwell-Chern-Simons
theory which was successfully constructed on the square lattice [15].
It turns out that the construction of an appropriate lattice action
with the correct continuum limit and direct evaluation of instanton
solutions are nontrivial because of the explicit
appearance of the gauge potential $a_{\mu}$ in the Chern-Simons term
and the fact that the gauge potential $a_{\mu}$ is not globally defined
in the presence of monopoles.
This problem was resolved by formulating the theory using the self-dual
model [15,21].
We will use the same idea and closely follow the derivations in Ref.
[15] to calculate the monopole-antimonopole action
from an equivalent self-dual model.

First of all, the equations of motion that are derived from the action
given in Eq.(3) is found to be
$$
\varepsilon ({\bf q}, \omega) \ q_{\alpha} \ e_{\alpha} ({\bf q}, \omega)
+ m \ b ({\bf q}, \omega) = 0 \ ,
$$
$$
\varepsilon ({\bf q}, \omega) \ \omega \ \epsilon_{\alpha \beta}
\ e_{\beta} ({\bf q}, \omega)
+ \mu ({\bf q}, \omega) \ q_{\alpha} \ b ({\bf q}, \omega)
- m \ e_{\alpha} ({\bf q}, \omega) = 0 \ ,
\eqno{(11)}
$$
where $\alpha = 1,2$.
Let us define $f_{\mu}$ as the dual of the field strength tensor
$f_{\mu \nu}$:
$f_{\mu} = \epsilon_{\mu \nu \lambda} f_{\nu \lambda}/2$ [15,21].
The Euclidean partition function of an equivalent dual theory can be
written as
$$
Z = \int \ D f_{\mu} \ D f^*_{\mu} \ e^{- S_{\rm E} (f_{\mu}, f^*_{\mu})} \ ,
$$
$$
\eqalign{
S_{\rm E} (f_{\mu}, f^*_{\mu})
&= \ {1 \over 2} \ \int {d^3 q \over (2 \pi)^3} \
\Bigl [ \ \mu ({\bf q}, \omega) \ f^*_0 \ f_0
- {1 \over m} \ \mu ({\bf q}, \omega) \ f^*_0 \
\varepsilon ({\bf q}, \omega) \ ( \ q_1 \ f_2
- q_2 \ f_1 \ ) \cr
&+ \varepsilon ({\bf q}, \omega) \ f^*_1 \ f_1
- {1 \over m} \ \varepsilon ({\bf q}, \omega) \ f^*_1 \
( \ \mu ({\bf q}, \omega) \ q_2 \ f_0
- \varepsilon ({\bf q}, \omega) \ \omega \ f_2 \ ) \cr
&+ \varepsilon ({\bf q}, \omega) \ f^*_2 \ f_2
- {1 \over m} \ \varepsilon ({\bf q}, \omega) \ f^*_2 \
( \ \varepsilon ({\bf q}, \omega) \ \omega \ f_1 -
\mu ({\bf q}, \omega) \ q_1 \ f_0 \ )
\ \Bigr ] \ ,
}
\eqno{(12)}
$$
where $f^*_{\mu} ({\bf q}, \omega) = f_{\mu} (-{\bf q}, -\omega)$.
One can easily check that the above action gives the same equations of motion
as Eq.(11).
In the lattice version of the action, as a result of the appropriate
regularization, we can separate out singularities from $f_{\mu}$ [15].
Now we can define the regularized dual field strength tensor $f^{reg}_{\mu}$
as $f_{\mu} = f^{reg}_{\mu} - {1 \over m} \ j_{\mu}$, where
the string singularity $j_{\mu}$ is given by Eq.(7).
This singularity acts as a source for $f^{reg}_{\mu}$ [15] and
the corresponding equations of motion for $f^{reg}_{\mu}$ can be written as
$$
\eqalign{
&f^{reg}_0 - {1 \over m} \
\varepsilon ({\bf q}, \omega) \ ( \ q_1 \ f^{reg}_2 - q_2 \ f^{reg}_1 \ )
= {1 \over m} \ j_0  \ , \cr
&f^{reg}_1 - {1 \over m} \
( \ \mu ({\bf q}, \omega) \ q_2 \ f^{reg}_0 - \varepsilon ({\bf q}, \omega) \
\omega \ f^{reg}_2 \ ) = 0 \ , \cr
&f^{reg}_2 - {1 \over m} \
( \ \varepsilon ({\bf q}, \omega) \ \omega \ f^{reg}_1 -
\mu ({\bf q}, \omega) \ q_1 \ f^{reg}_0 \ ) = 0 \ . }
\eqno{(13)}
$$
Inverting these equations, we can get the following solutions,
$$
f^{reg}_{\mu} ({\bf q}, \omega)
= {1 \over m} \ G_{\mu \nu} ({\bf q}, \omega) \
j_{\nu} ({\bf q}, \omega) \ ,
\eqno{(14)}
$$
where
$$
\eqalign{
G_{\mu \nu} ({\bf q}, \omega) &= (A^{-1})_{\mu \nu} ({\bf q}, \omega) \ , \cr
A_{\mu \nu} ({\bf q}, \omega) &=
\pmatrix{ 1 & q_2 \ \varepsilon ({\bf q}, \omega) / m
& - q_1 \ \varepsilon ({\bf q}, \omega) / m \cr
- q_2 \ \mu ({\bf q}, \omega) / m & 1
& \omega \ \varepsilon ({\bf q}, \omega) / m \cr
q_1 \ \mu ({\bf q}, \omega) \ / m
& - \omega \ \varepsilon ({\bf q}, \omega) / m & 1 } \ . }
\eqno{(15)}
$$
The Eq.(14) represents the monopole-antimonopole solution of the
effective gauge theory.

The monopole-antimonopole action can be obtained from Eqs.(12) and (14) [15].
$$
\eqalign{
S_{M {\bar M}} (\tau) &= {1 \over 2} \int {d^3 q \over (2 \pi)^3}
\ {1 \over m^2} \
\ j_{0} (-{\bf q}, - \omega) \ \mu ({\bf q}, \omega) \ G_{00} ({\bf q}, \omega)
\ j_{0} ({\bf q}, \omega) \cr
&= {1 \over 2} \int {d^3 q \over (2 \pi)^3} \ {1 \over m^2} \
\ j_{0} (-{\bf q}, - \omega)
\ { (m^2 + \varepsilon^2 ({\bf q}, \omega) \ \omega^2) \ \mu ({\bf q}, \omega)
\over m^2 + \varepsilon ({\bf q}, \omega) \ (\varepsilon ({\bf q}, \omega) \
\omega^2 + \mu ({\bf q}, \omega) \ q^2) }
\ j_{0} ({\bf q}, \omega) \ .
}
\eqno{(16)}
$$
Notice that the appearance of $m^2$ in the fractional expression
reflects the screening effect of the
Chern-Simons term. In the Maxwell-Chern-Simons theory, $\varepsilon = \mu =1$
so that $m^2$ term dominates in the infrared limit and this screening effect
confines the monopole-antimonopole pairs [15].
However, in our model, $\varepsilon$
and $\mu$ are divergent ( $\mu$ is divergent for $\eta > 0$ )
in the infrared limit so that $m^2$ term becomes
irrelevant. Therefore, the contributions from the dielectric function
and the magnetic permeability due to the particle-hole excitations
dominate the screening effect of the Chern-Simons term.
Now we can safely set $m^2 = 0$ in the numerator and the denominator of the
fractional expression in Eq.(16), then $S_{M {\bar M}} (\tau)$ can
be written as
$$
S_{M {\bar M}} (\tau) = \int {d^2 q \over (2 \pi)^2}
\int {d \omega \over 2 \pi} {(2 \pi {\tilde \phi})^2 \over \omega^2}
{\varepsilon ({\bf q}, \omega) \ \mu ({\bf q}, \omega) \ \omega^2 \over
\varepsilon ({\bf q}, \omega) \ \omega^2 + \mu ({\bf q}, \omega) \ q^2}
(1 - {\rm cos}(\omega \tau)) \ ,
\eqno{(17)}
$$
where ${\tilde \phi} = 2$ for $\nu = 1/2$. The above result can be
easily understood once we realize that the Chern-Simons term is irrelevant
in the infrared limit. After dropping the Chern-Simons term, the
effective action $S_{\rm eff}(a_\mu)$ (see Eq.(3))
is essentially the Maxwell theory
with frequency and momentum dependent dielectric function
$\varepsilon ({\bf q}, \omega)$ and
magnetic permeability $\mu ({\bf q}, \omega)$. (17) is just the
monopole-antimonopole action in this generalized Maxwell theory [16].

For the Coulomb interaction $v({\bf q}) = {2 \pi e^2 \over \epsilon q}$,
$\mu$ can be approximated as
${2 \pi e^2 \over \epsilon (2 \pi {\tilde \phi})^2}{1 \over q}$:
$$
\eqalign{
S_{M {\bar M}} (\tau) &=
{e^2 \over 2 \pi \epsilon} \sqrt{\tau \over \beta} \int^{\infty}_{0} dx
\int^{\infty}_{0} dy {1 \over x y^{3/2}}{1 \over 1 + x}(1-{\rm cos}(xy))
\ , \cr
&\equiv 2 \sqrt{\Omega_0 (1) \tau} \ ,
}
\eqno{(18)}
$$
where $\beta = e^2 l_c / 4 \epsilon$ ($l_c$ is the magnetic length) [11] and
$\Omega_0 (1) = \pi e^2 / \epsilon l_c$.
Therefore, the monopole-antimonopole pair is confining but the action
is proportional to the square-root of the distance between monopole and
antimonopole which is
different from the linearly confining monopole-antimonopole solution of
the Maxwell-Chern-Simons theory. The confinement of monopoles means the
existence of a well defined `charged' particle or composite fermion.

The same calculation can be done for a class of interactions $v({\bf q}) =
V_0 / q^{\eta} \ ( 0 \leq \eta < 2 )$.
Using the fact that
$\mu \approx {V_0 \over (2 \pi {\tilde \phi})^2} {1 \over q^{\eta}} \
( \eta > 0 )$ and $\mu = {1 \over 12 \pi m^*} +
{V_0 \over (2 \pi {\tilde \phi})^2} \ ( \eta = 0 )$, we get
$S_{M {\bar M}} (\tau) =
(3 - \eta) \left [ \Omega_0 (\eta) \ \tau \right ]^{1 \over (3-\eta)}$
with
$$
\eqalign{
\Omega_0 (\eta \not= 0) &= {V_0 l_c \over 4 \pi} \ \left [
{2 \pi \over (3 - \eta)^2 l_c} \
{1 \over \Gamma ({4 - \eta \over 3 - \eta})} \
{\rm cosec} \left ( {\pi \over 2 (3 - \eta)} \right ) \
{\rm cosec} \left ( {\pi \over (3 - \eta)} \right )
\right ]^{3 - \eta} \ , \cr
\Omega_0 (\eta = 0) &= {2^{11} \pi^4 \over 3^{17/2} \Gamma^3 (4/3)} \
{{\tilde \chi} \over l^2_c} \ ,
}
\eqno{(19)}
$$
where ${\tilde \chi} = {1 \over 12 \pi m^*} +
{V_0 \over (2 \pi {\tilde \phi})^2}$ is the effective diamagnetic
susceptibility of the fermions.
Therefore, the monopole-antimonopole pair is still confining.

{}From (10) and (19), we can see that the
electron Green's function has a form
$G_{+} (\tau) \approx G_0 (\tau)
e^{(3 - \eta) \left [ \Omega_0 (\eta) \ \tau \right ]^{1 \over (3-\eta)}}$.
After relatively unimportant factor $G_0$ is dropped, it has
the same functional form as the result of He, Platzman, and Halperin [11]
in the case of the Coulomb interaction ($\eta = 1$).
Notice that our $\Omega_0 (1)$ is two-times larger than $\omega_0$ they
obtained in a similar expression [11].
We found that, for $0 \leq \eta < 2$, the constant $\Omega_0 (\eta)$ has
different functional dependence on $\eta$ compared to the result
one would get if the approach [11] of X-ray edge problem was used.
In X-ray edge problem, the added electron, being a core electron,
does not participate in density fluctuations of the valence electrons.
In our problem, the added electron, being a valence electron,
contributes to the density fluctuations (through the flux it carries).
Thus one should take into account properly the flux of the added electron
which is an important part of the gauge field configuration.
In our instanton approach, the flux degree of freedom of the
added electron is naturally included.
The low frequency behavior of the corresponding
spectral function $A_{+} (\omega)$, which is the inverse Laplace
transform of $G_{+}$ [11], is given by
$e^{ - (\Omega_0 (\eta) / \omega)^{1 \over (2-\eta)} }$.
It was pointed out that the exponential suppression of the spectral
density leads to the strong suppression of the tunneling current
at low voltage biases [11]. These results show that
the one electron Green's function has very different behavior compared
to that in the usual Fermi liquid although two-particle correlation functions
may be Fermi-liquid-like [18,22].

Recently, Bonesteel [23] extended the analysis of Ref.[4] to the double-layer
system near $\nu = 1/2$. It was found that
the dynamics of the  gauge
field fluctuations in two layers separates into out-of-phase mode
and in-phase mode between two layers.
The out-of-phase mode behaves as if there is no Coulomb interaction
[23]. The tunneling between two layers corresponds to the creation of
a monopole in one layer and an antimonopole in the other
layer which only couple to the out-of-phase mode of the gauge field.
Thus the tunneling current will be directly proportional to
$e^{- S_{M {\bar M}} (\tau)}$ with $\eta = 0$ (for short range interaction).
Replacing ${\tilde \chi}$ by
the appropriate effective diamagnetic susceptibility [23] of
out-of-phase current fluctuations, we get
$I(V) \sim e^{- (8 \Omega_0 (0) / e V)^{1/2} }$ where
the factor $8$ comes from the existence of two layers. We expect the
above to be valid at low biases where the interlayer screening becomes
important.

In summary, we investigated the instanton-anti-instanton or
monopole-antimonopole solution of the effective gauge theory of
the half-filled Landau level.
It was found that instanton-anti-instanton pairs are confining so that
a well defined composite fermion can exist.
We also related the instanton-anti-instanton action to the electron Green's
function in the semiclassical approximation.
It was found that the strong suppression of the spectral function in the
infrared limit can be understood in terms of the Euclidean-time dependence
of the instanton-anti-instanton action.

Acknowledgement: The authors thank Dongsu Bak, B. I. Halperin,
P. A. Lee, N. Nagaosa, and P. C. E. Stamp for the discussions about
this and related issues.
This work was supported by NSF grant No. DMR-91-14553.

\vfill\vfill\vfill
\break

\vskip 0.5cm

\centerline{\bigbf References}

\vskip 0.5cm

\item{[1]} J. K. Jain, Phys. Rev. Lett. {\bf 63}, 199 (1989);
Phys. Rev. {\bf B} {\bf 41}, 7653 (1990); Adv. Phys. {\bf 41}, 105 (1992).
\item{[2]} A. Lopez and E. Fradkin, Phys. Rev. {\bf B} {\bf 44}, 5246 (1991).
\item{[3]} V. Kalmeyer and S. C. Zhang, Phys. Rev. {\bf B} {\bf 46},
9889 (1991).
\item{[4]} B. I. Halperin, P. A. Lee, and N. Read,
Phys. Rev. {\bf B} {\bf 47}, 7312 (1993).
\item{[5]} R. L. Willet, M. A. Paalanen, R. R. Ruel, K. W. West,
L. N. Pfeiffer, and D. J. Bishop, Phys. Rev. Lett. {\bf 65}, 112 (1990).
\item{[6]} R. L. Willet, R. R. Ruel, M. A. Paalanen, K. W. West,
and L. N. Pfeiffer, Phys. Rev. {\bf B} {\bf 47}, 7344 (1993).
\item{[7]} R. R. Du, H. L. Stormer, D. C. Tsui, L. N. Pfeiffer,
and K. W. West, Phys. Rev. Lett. {\bf 70}, 2944 (1993).
\item{[8]} W. Kang, H. L. Stormer, L. N. Pfeiffer, K. W. Baldwin,
and K. W. West, Phys. Rev. Lett. {\bf 71}, 3850 (1993).
\item{[9]} A. M. Polyakov, Phys. Lett. {\bf 59 B}, 82 (1975);
Nucl. Phys. {\bf B} {\bf 120}, 429 (1977).
\item{[10]} J. P. Eisenstein, L. N. Pfeiffer, and K. W. West,
Phys. Rev. Lett. {\bf 69}, 3804 (1992).
\item{[11]} S. He, P. M. Platzman, and B. I. Halperin,
Phys. Rev. Lett. {\bf 70}, 777 (1993).
\item{[12]} Y. Hatsugai, P.-A. Bares, and X.-G. Wen,
Phys. Rev. Lett. {\bf 71}, 424 (1993).
\item{[13]} P. D. Pisarski, Phys. Rev. {\bf D} {\bf 34}, 3851 (1986).
\item{[14]} I. Affleck, J. Harvey, L. Palla, and G. Semenoff, Nucl. Phys.
{\bf B 328}, 575 (1989).
\item{[15]} M. C. Diamantini, P. Sodano, and C. A. Trugenberger,
Phys. Rev. Lett. {\bf 71}, 1969 (1993).
\item{[16]} L. B. Ioffe and A. I. Larkin, Phys. Rev. {\bf B} {\bf 39},
8988 (1989).
\item{[17]} N. Nagaosa, Phys. Rev. Lett. {\bf 71}, 4210 (1993).
\item{[18]} Junwu Gan and Eugene Wong, Phys. Rev. Lett. {\bf 71},
4226 (1994); J. Polchinski, UCSB preprint.
\item{[19]} R. Rajaraman, {\it Solitons and Instantons}.
(North-Holland, Amsterdam, Netherland, 1982).
\item{[20]} B. L. Altshuler and L. B. Ioffe, Phys. Rev. Lett. {\bf 69},
2979 (1992).
\item{[21]} S. Deser and R. Jackiw, Phys. Lett. {\bf 139 B}, 371 (1984).
\item{[22]} D. V. Khveshchenko and P. C. E. Stamp, Phys. Rev. Lett. {\bf 71},
2118 (1993); Phys. Rev. {\bf B} {\bf 49}, 5227 (1994).
\item{[23]} N. Bonesteel, Phys. Rev. {\bf B} {\bf 48},
11484 (1993): The effective diamagnetic susceptibility of the out-of-phase
current fluctuation is given by ${1 \over 12 \pi m^*} +
{(1 + e^2 m^* d) \over 2 \pi
{\tilde \phi}^2 m^*}$ where $d$ is the inter-layer distance.

\bye